\documentstyle[12pt]{article}

\parskip=\medskipamount 
\def\be{\begin{equation}}
\def\bea{\begin{eqnarray}}
\def\b*{\begin{eqnarray*}}
\def\nn{\nonumber}
\def\ee{\end{equation}}
\def\eea{\end{eqnarray}}
\def\e*{\end{eqnarray*}}

\begin{document}
\begin{titlepage}
\begin{flushright}
hep-th 9401168
\end{flushright}

\vspace{2cm}

\begin{center}
{\Large \bf The Number Operator for Generalized Quons}\\
\vspace{2.cm}
{\normalsize \bf Miroslav Dore\v si\' c \footnote
{e-mail address: doresic@thphys.irb.hr }}\\

\vspace{0.5cm}

Department of Theoretical Physics, \\
Rudjer Bo\v skovi\' c Institute, P.O.B. 1016,\\
41001 Zagreb, CROATIA \\
\vspace{2cm}
{\large \bf Abstract}
\end{center}
\vspace{0.5cm}
\baselineskip=24pt

We construct the number operator for particles obeying
infinite statistics,\\ defined
by a generalized q-deformation of the Heisenberg algebra,
and prove the \\
positivity of the norm of linearly independent
state vectors.

\vspace{50mm}

\begin{center}
\end{center}
\end{titlepage}

\setcounter{page}{1}
\newpage
\baselineskip=24pt
\vspace{1.5cm}     

The approach to particle statistics based on deformations
of the bilinear
Bose and Fermi commutation relations has attracted considerable
interest during the last few years \cite{{gr1},{gr2},{moh},{fiv}}.
The particles obeying this type of
statistics are called "quons". The quon algebra (or the q$-$mutator)
is given by
\be
a_{i} a_j^{\dagger} - q a_j^{\dagger} a_i = {\delta}_{ij},
\;\;\; (\forall i,j),                                       \label{1}
\ee
and interpolates between Bose and Fermi algebras as the deformation
parameter $q$ goes from 1 to $-1$ on the real axis. When
supplemented by the vacuum condition
\be
a_i \mid 0 > = 0,\;\;\; (\forall i),                        \label{2}
\ee
the quon algebra determines a (Fock-like) representation in
a linear vector space. For $q \in $[-1,1], the squared norms
of all vectors made by the limits
of the polynomials of the creation operators~$a_k^{\dagger}$~
are strictly positive. No commutation relation can be imposed
on ~$a_i a_j$~ or ~$a_i^{\dagger} a_j^{\dagger}$. Furthermore,
no such rule is needed
to calculate the vacuum matrix elements of
the polynomials in the~$a$'s~
and~$a^{\dagger}$'s. All such matrix elements can be calculated
by moving the annihilation operators to the right using (\ref{1}),
until they, according to (\ref{2}), annihilate the vacuum
\cite{gr}.

The aim of this paper is to construct the number operator for a
generalized $q-$deformation of the Heisenberg algebra which is
characterized by the following relations
\bea
a_{i} a_{j}^{\dagger} - q_{ij} a_{j}^{\dagger} a_{i} & = &
{\delta}_{ij}, \;\;\; (\forall i,j ),                       \label{3}
\\
a_{i} \mid 0 > &=& 0  \; \; \;( \forall i ),                \label{4}
\\
q_{ji}^{\ast} &=& q_{ij},                                   \label{5}
\eea
with the deformations parameters $q_{ij}$ being, in general, complex
numbers. The statistics based on the commutation relations (\ref{3})
generalize classical Bose and Fermi statistics, which correspond to
{}~$q_{ij} = 1 \; ( \forall i,j)$~ and
{}~$q_{ij} = -1 \; ( \forall i,j)$, respectively.
The relation (\ref{5}) follows from the consistency requirement of
the relation (\ref{3}).

For each $k$, the $k^{th}$ number operator $N_k$ satisfies the
relations
\bea
N_{k}^{\dagger} = N_{k} \; ,& & \; \; \; N_{k} \, \mid 0 > = 0 ,
\nn \\
{[} N_{k} , a_{l}^{\dagger} ] & = & {\delta}_{kl} a_{l}^{\dagger}
\; \; \; \; \; \; \; ( \forall l ) \; ,                     \label{6}
\eea
which is equivalent to
\be
N_{k} \: ( a_{i_{1}}^{\dagger} a_{i_{2}}^{\dagger}
\cdots a_{i_{n}}^{\dagger} \mid 0 > ) = s \:
( a_{i_{1}}^{\dagger} a_{i_{2}}^{\dagger}
\cdots a_{i_{n}}^{\dagger} \mid 0 > ) \;,                   \label{7}
\ee
where $s$ is the index number $i_{j}$, such that $i_{j}=k$.
The most general expression for the number operator $N_k$ is
of the form
\bea
N_{k} & = & a_{k}^{\dagger} a_{k} + \left.\sum_{n=2}^{\infty}\right.
\left.\sum_{({\bf i}_{n-1})}\right.
\left.\sum_{\pi (k,{\bf i}_{n-1})}\right.
\left.\sum_{\sigma (k,{\bf i}_{n-1})}\right.
c_{\pi (k,{\bf i}_{n-1}) , \sigma (k,{\bf i}_{n-1})}
\nn \\
& & \; \cdot \; a_{\pi (k)}^{\dagger}a_{\pi ( i_{1})}^{\dagger} \cdots
a_{\pi ( i_{n-1})}^{\dagger}
a_{\sigma (k)} a_{\sigma ( i_{1})} \cdots
a_{\sigma ( i_{n-1})} \; ,                                  \label{8}
\eea
where ~${\bf i}_{n-1} \equiv (i_{1}, \dots , i_{n-1} )$ is an
arbitrary choice of n$-$1 indices, {\bf including their repetitions},
whereas ~$\pi , \sigma $~are permutations of n indices.
Making use of the relation (\ref{7}), one finds that the condition
satisfied by the coefficients
{}~$c_{i_{1}, \dots , i_{n} ;j_{1}, \dots , j_{n}}$~
can be written in the form of the following matrix equation
\be
[ ( c_{i_{1}, \dots , i_{n} ;j_{1}, \dots , j_{n}} ) ] \; \times \;
M_{n}(q) \;  =  \; Q_{n}(q) \; .                            \label{9}
\ee
The matrix $M_{n}$ is defined by
\be
( M_{n} )_{i_{1}, \dots , i_{n} ;j_{1}, \dots , j_{n}} \; = \;
< 0 \mid a_{i_{n}} \cdots a_{i_{1}} a_{j_{1}}^{\dagger} \cdots
a_{j_{n}}^{\dagger} \mid 0 > \; .                           \label{10}
\ee
Taking into account eqs. (\ref{3}), (\ref{4}), (\ref{5}) and
(\ref{10}), and
using the method of matemathical induction, one arrives at the
closed-form expression
\be
(M_{n} )_{\pi (1, \dots , n ) ; \sigma (1, \dots , n )} =
\left.\prod_{r,s=1}^{n}\right. q_{\pi (r) \sigma (s)}^{P(r,s)},
\;\; ( \pi (r) \neq \sigma (s) ),                           \label{11}
\ee
where
\be
P(r,s)=\theta [ - ( r - s ) (({\sigma}^{-1}
\cdot \pi)(r) - ({\sigma}^{-1} \cdot \pi )(s))],            \label{12}
\ee
and $\theta (x)$ designates the function defined by
\be
\theta (x) = \left\{ \begin{array}{ll}
                     1       & {\rm if} \; x > 0 \: , \\
                     0       & {\rm if} \; x \leq 0 \: .
                     \end{array}
             \right.                                        \label{13}
\ee
The determinant of the matrix $M_n$ is
\bea
\det M_{n} & = & \left.\prod_{k=1}^{n-1}\right.
[ \left.\prod_{\{ i_{1}, \dots , i_{k+1} \}}\right.
\, ( \, 1- \left.\prod_{\{ i_{\alpha} i_{\beta} \}}\right.
\mid q_{i_{\alpha} i_{\beta}} {\mid}^2 \, ) \; ]^{(k-1)!(n-k)!}
\; ,                                                        \label{14}
\eea
where the set $\{ i_{1}, \dots , i_{k+1} \}$ denotes a choice of $k+1$
different indices out of $n$ such indices,whereas
$\{ i_{\alpha} i_{\beta} \}$ is any of its subsets.
Unlike in the case of the matrix $M_n(q)$, explicitly given by eqs.
(\ref{11}) and (\ref{12}), the closed-form expression for
the matrix $Q_n(q)$ cannot be written down. In fact, this matrix
is obtained as a result of the action
of the lower-order terms
$a_{i_{1}}^{\dagger} \cdots a_{i_{s}}^{\dagger}
a_{j_{1}} \cdots a_{j_{s}}, \; s<n$, entering expression (\ref{8})
for the number operator on the eigenvector
$a_{k_{1}}^{\dagger} \cdots a_{k_{n}}^{\dagger} \mid 0 >$. Thus, to
completely determine the matrix $Q_n(q)$, knowledge of all
the lower-order coefficients
$c_{i_{1}, \dots , i_{s} ;j_{1}, \dots , j_{s}}$,\\
$s=2,3,\dots, n-1$ is required.

As a special case of the general formula (\ref{14}),
we give the expression
for the determinant of the matrix $M_4$, corresponding to n=4
and $(k_{1},k_{2},k_{3},k_{4}),
\equiv (k,l,m,p)$:
\bea
\det M_{4} &=& (1-\mid q_{kl}{\mid}^2 )^6 \;
(1-\mid q_{km}{\mid}^2 )^6
\; (1-\mid q_{kp}{\mid}^2 )^6 \; (1-\mid q_{lm}{\mid}^2 )^6 \;
\nn \\
& & \times (1-\mid q_{lp}{\mid}^2 )^6
\; (1-\mid q_{mp}{\mid}^2 )^6
\nn \\
& & \times \; (1-\mid q_{kl}{\mid}^2 \mid q_{km}{\mid}^2
\mid q_{lm}{\mid}^2 )^2 \;
(1-\mid q_{kl}{\mid}^2 \mid q_{kp}{\mid}^2
\mid q_{lp}{\mid}^2 )^2
\nn \\
& & \times \; (1-\mid q_{km}{\mid}^2 \mid q_{kp}{\mid}^2
\mid q_{lp}{\mid}^2 )^2 \;
(1-\mid q_{lm}{\mid}^2 \mid q_{lp}{\mid}^2
\mid q_{mp}{\mid}^2 )^2
\nn \\
& & \times \; (1-\mid q_{kl}{\mid}^2 \mid q_{km}{\mid}^2
\mid q_{kp}{\mid}^2 \mid q_{lm}{\mid}^2 \mid q_{lp}{\mid}^2
\mid q_{mp}{\mid}^2 ).                                      \label{15}
\eea
It is evident from eq.(\ref{14}), and especially from the special
case (\ref{15}), that if the \\
deformation parameters are such that
$\mid q_{ij} \mid < 1 \; \; ( \forall i,j )$,
the matrix $M_n(q)$ is regular and positively definite.
Consequently, the coefficients
$c_{i_{1}, \dots , i_{n} ;j_{1}, \dots , j_{n}}$, appearing
in expression (\ref{8}) for the number operator $N_k$, exist
and the norm
of the state vector $ a_{i_{1}}^{\dagger}
a_{i_{2}}^{\dagger} \cdots a_{i_{n}}^{\dagger} \mid 0 >$ is positive.
That being the case, one is now allowed to rewrite \\
eq. (\ref{9}) in the form
\be
C_n(q) = Q_{n}(q) \; \times \; [M_{n}(q)]^{-1}\; ,           \label{16}
\ee
where, for notational simplicity,
\be
C_n(q) = [ ( c_{i_{1}, \dots , i_{n} ;j_{1}, \dots ,
j_{n}} ) ] \; .                                             \label{17}
\ee
Equation (\ref{16}) represents the main result of this paper.

As an example, in the following we illustrate the calculation of the
coefficients $ c_{i_{1}, i_{2}, i_{3} ;j_{1}, j_{2}, j_{3}}$, which
amounts to finding the matrix coefficients $C_3(q)$. According to
(\ref{16}), this matrix is given by
\be
C_3(q) = Q_{3}(q) \; \times \; [M_{3}(q)]^{-1}\;.            \label{18}
\ee
There are four different cases to be considered.

\underline{{\bf Case 1:}} ~~~ n=3
$(k_{1},k_{2},k_{3})\equiv (k,k,k)$. \\
This case is trivial, since $Q_3(q)$ and $M_3(q)$ are represented by
the numbers (1$-$square matrices)
\be
{(Q_3(q))}_{kkk,kkk} = (1- q_{kk})^2(1+ q_{kk}),            \label{19}
\ee
\be
{(M_3(q))}_{kkk,kkk} = (1+ q_{kk})(1+ q_{kk} + q_{kk}^2),   \label{20}
\ee
so that, in view of (\ref{18}),
\be
c_{kkk,kkk} = \frac{(1-  q_{kk} )^2}{1+ q_{kk} +q_{kk}^2} . \label{21}
\ee

\underline {{\bf Case 2:}} ~~~ n=3
$(k_{1},k_{2},k_{3})\equiv (k,k,l)$.\\
In this case, $Q_3(q)$ and $M_3(q)$ are 3$-$square matrices,
the rows and columns of which
are indexed in the order (k,k,l), (k,l,k) and (l,k,k).
The matrix $M_3(q)$ is given by
\be
{M}_{3}(q)  =
    \left[
    \begin{array}{ccc}
1+q_{kk} & q_{kl}(1+q_{kk}) & q_{kl}^2 (1+q_{kk}) \\
q_{lk}(1+q_{kk}) & 1+ q_{kk} \mid q_{kl}{\mid}^2 & q_{kl}(1+q_{kk})\\
q_{lk}^2 (1+q_{kk}) & q_{lk} (1+q_{kk}) & (1+q_{kk})
    \end{array}
    \right] \; ,                                            \label{22}
    \ee
with the determinant
\be
\det M_3 \: = \: (1+ q_{kk})^2 \; (1-\mid q_{kl}{\mid}^2 )^2
\; (1- q_{kk} \mid q_{kl}{\mid}^2 ) \, .                    \label{23}
\ee
On the basis of eqs. (\ref{22}) and (\ref{23}), the
inverse matrix of $M_3(q)$ is found to be
\bea
[{M}_{3}(q)]^{-1} & = & \frac{1}{(1+q_{kk})
                               (1-q_{kk} \mid q_{kl} {\mid}^2 )}
\nn \\
& & \times
\left[
    \begin{array}{ccc}
1 & - q_{kl}(1+q_{kk}) & q_{kk} q_{kl}^2  \\
- q_{lk}(1+q_{kk}) & (1+ q_{kk})(1+ \mid q_{kl}{\mid}^2 )
 & - q_{kl}(1+q_{kk})\\
q_{kk} q_{lk}^2 &- q_{lk} (1+q_{kk}) & 1
    \end{array}
    \right] \; .                                            \label{24}
\eea
The matrix $Q_3$ can be obtained using the following lower-order
coefficients:
\bea
& & \hspace{1.7cm} c_{kk,kk}  =  \frac{1-  q_{kk}}{1+ q_{kk}},
\nn \\
c_{kl,kl} & = & - \frac{q_{kl}}{1- \mid q_{kl} {\mid}^2} ,
\hspace{2cm}
c_{lk,kl}  =  \frac{1}{1- \mid q_{kl} {\mid}^2} ,
\nn \\
c_{kl,lk} & = & \frac{\mid q_{kl} {\mid}^2}{1- \mid q_{kl} {\mid}^2} ,
\hspace{2.3cm}
c_{lk,kl}  =  \frac{-  q_{lk}}{1- \mid q_{kl} {\mid}^2} .   \label{25}
\eea
The result is
\bea
Q_{3}(q) & = &
    \left[
    \begin{array}{ccc}
0 & - q_{kl} & - q_{kl}^2 (1- q_{kk}) \\
0 & 1+ q_{kk} \mid q_{kl}{\mid}^2 & 0 \\
0 & - q_{kk} q_{lk} & -1+q_{kk}
    \end{array}
        \right] \; .                                            \label{26}
\eea
The coefficient matrix $C_3(q)$ is now obtained by substituting
eqs. (\ref{24}) and (\ref{26}) into eq.(\ref{18}).
To see what the structure of these
coefficients is like, we only give the elements of the
first row of this matrix:
\bea
    \begin{array}{ll}
c_{kkl,kkl} = 2 q_{kk} q_{kl}^2 \cdot \Delta , &  \\
c_{kkl,klk} = - q_{kl}(1+q_{kk})
              (1 + q_{kk} \mid q_{kl} {\mid}^2 ) \cdot \Delta , & \\
c_{kkl,lkk} = \mid q_{kl} {\mid}^2
                  ( 1 + q_{kk} - q_{kk} \mid q_{kl} {\mid}^2
 + {q_{kk}}^2 \mid q_{kl}{\mid}^2 ) \cdot \Delta , &
    \end{array}                                             \label{27}
\eea
with
\be
\Delta =  \frac{1}{(1+q_{kk})
                   (1-q_{kk} \mid q_{kl} {\mid}^2 )} \, .   \label{28}
\ee

\underline {{\bf Case 3:}} ~~~ n=3
$(k_{1},k_{2},k_{3})\equiv (k,k,l)$.\\
In this case,the results are of the same structure as those
in the preceding case, so we do not give them here.

\underline {{\bf Case 4:}} ~~~ n=3
$(k_{1},k_{2},k_{3})\equiv (k,l,m)$.\\
In this case, $Q_3(q)$ and $M_3(q)$ are 6$-$square matrices,
the rows and columns of which are indexed by the elements of the
permutation group in the order (k,l,m), (l,k,m), (k,m,l),
(l,m,k), (m,k,l) and (m,l,k). Again, the matrix
$Q_3(q)$ can be obtained with the help of the lower-order
coefficients (\ref{25}), and is given by
\bea
Q_{3}(q) & = &
    \left[
    \begin{array}{cccccc}
q_{kl}q_{km}q_{lm} & 0 & 0 & 0 & 0 & 0 \\
- q_{km}q_{lm} & 0 & - q_{km} & 0 & 0 & 0 \\
0 & 0 & q_{kl}q_{km}q_{ml} & 0 & 0 & 0 \\
0 & 0 & 1 & 0 & 0 & 0 \\
- q_{kl} & 0 & - q_{kl}q_{ml} & 0 & 0 & 0 \\
1 & 0 & 0 & 0 & 0 & 0
    \end{array}
    \right]     \; .                                        \label{29}
\eea
On the basis of eq. (\ref{11}), the matrix $M_3(q)$ is found to be
\be
M_{3}(q)  =
    \left[
    \begin{array}{cccccc}
1 & q_{kl} & q_{lm} & q_{kl}q_{km} & q_{km}q_{lm} & q_{kl}q_{km}q_{lm}\\
q_{lk} & 1 & q_{lk}q_{lm} & q_{km} & q_{lm}q_{lk}q_{km} & q_{lm}q_{km}\\
q_{ml} & q_{kl}q_{ml} & 1 & q_{kl}q_{km}q_{ml} & q_{km} & q_{km}q_{kl}\\
q_{lk}q_{mk} & q_{mk} & q_{lk}q_{lm}q_{mk} & 1 & q_{lm}q_{lk} & q_{lm}\\
q_{mk}q_{ml} & q_{ml}q_{mk}q_{kl} & q_{mk} & q_{ml}q_{kl} & 1 & q_{kl}\\
q_{mk}q_{ml}q_{lk} & q_{ml}q_{mk} & q_{mk}q_{lk} & q_{ml} & q_{lk} & 1
    \end{array}
    \right]  \; ,                                           \label{30}
\ee
with the determinant
\bea
\det M_{3}(q) &=& (1-\mid q_{kl}{\mid}^2 )^2 \;
(1-\mid q_{km}{\mid}^2 )^2
\; (1-\mid q_{lm}{\mid}^2 )^2
\nn \\
& & \times (1-\mid q_{kl}{\mid}^2 \mid q_{km}{\mid}^2
\mid q_{lm}{\mid}^2 ).                                       \label{31}
\eea
Inserting the matrices $Q_3(q)$, given by eq.(\ref{29}),
and the matrix
$M_3^{-1}(q)$, obtainable from eqs.(\ref{30}) and (\ref{31})),
into eq.(\ref{18}),
one finds the coefficient matrix $C_3(q)$. Here, as in the
{\bf Case 2}, we only exibit the elements comprising its first row.
They are
\bea
c_{klm,klm} & = & (M_3^{-1})_{klm,mlk},
\nn \\
c_{klm,lkm} & = & (M_3^{-1})_{lkm,mlk},
\nn \\
c_{klm,kml} & = & (M_3^{-1})_{kml,mlk},
\\
c_{klm,lmk} & = & (M_3^{-1})_{lmk,mlk} \mid q_{kl} {\mid}^2
              \mid q_{km} {\mid}^2\mid q_{lm} {\mid}^2,
\nn \\
c_{klm,mkl} & = & (M_3^{-1})_{mkl,mlk} \mid q_{kl} {\mid}^2
              \mid q_{km} {\mid}^2\mid q_{lm} {\mid}^2,
\nn \\
c_{klm,mlk} & = & (M_3^{-1})_{mlk,mlk} \mid q_{kl} {\mid}^2
              \mid q_{km} {\mid}^2\mid q_{lm} {\mid}^2,
\nn                                                         \label{32}
\eea
where
\bea
(M_3^{-1})_{klm,mlk} & = & q_{kl}q_{km}q_{lm}
(1- \mid q_{km} {\mid}^2 )
(1- \mid q_{kl}{\mid}^2 \mid q_{lm} {\mid}^2 ),
\nn \\
(M_3^{-1})_{lkm,mlk} & = & - \mid q_{kl} {\mid}^2 q_{km}q_{lm}
(1- \mid q_{km} {\mid}^2 )
(1- \mid q_{lm} {\mid}^2 ),
\nn \\
(M_3^{-1})_{kml,mlk} & = & - q_{kl}q_{km} \mid q_{lm}{\mid}^2
(1- \mid q_{kl} {\mid}^2 )
(1- \mid q_{km} {\mid}^2 ),
\\
(M_3^{-1})_{lmk,mlk} & = & - q_{lm}
(1- \mid q_{kl} {\mid}^2 )
(1- \mid q_{km} {\mid}^2 ),
\nn \\
(M_3^{-1})_{mkl,mlk} & = & - q_{kl}
(1- \mid q_{km} {\mid}^2 )
(1- \mid q_{lm} {\mid}^2 ),
\nn \\
(M_3^{-1})_{mlk,mlk} & = &
(1- \mid q_{km} {\mid}^2 )
(1- \mid q_{kl}{\mid}^2 \mid q_{lm} {\mid}^2 ),
\nn                                                         \label{33}
\eea
are the relevant matrix elements of $M_3^{-1}(q)$.

Next we derive a relation that will make it possible to rewrite
expression (\ref{8}), for the number operator $N_k$,
in a more compact and elegant form.

Making use of eqs. (\ref{11}) and (\ref{25}),
one finds that, for the case
$n=2$ and $(k_{1},k_{2})\equiv (k,l)$, the following relation holds:
\be
\left.\sum_{\pi (k,l)}\right.
\left.\sum_{\sigma (k,l)}\right.
c_{\pi (k,l) , \sigma (k,l)}
a_{\pi (k)}^{\dagger}a_{\pi ( l)}^{\dagger}
a_{\sigma (k)} a_{\sigma (l)}  =
( M_2^{-1} )_{kl,kl} {\tilde a}_{kl}^{\dagger}{\tilde a}_{kl} \; .
                                                            \label{34}
\ee
For the case $n=3$ and
$(k_{1},k_{2},k_{3})\equiv (k,l,m)$, an analogous relation
can be established
on the basis of the eqs. (\ref{32}) i (\ref{33}). It reads
\bea
\left.\sum_{\pi (k,l,m)}\right.
\left.\sum_{\sigma (k,l,m)}\right.
c_{\pi (k,l,m) , \sigma (k,l,m)}
a_{\pi (k)}^{\dagger}a_{\pi (l)}^{\dagger}a_{\pi (m)}^{\dagger}
a_{\sigma (k)} a_{\sigma (l)}a_{\sigma (m)}  &  &
\nn \\
= \; \left.\sum_{\pi (l,m)}\right.
\left.\sum_{\sigma (l,m)}\right.
( M_3^{-1} )_{k,\sigma (l,m) ; k,\pi (l,m)}
{\tilde a}_{k,\pi (l,m)}^{\dagger}{\tilde a}_{k, \sigma (l,m)}, & &
                                                            \label{35}
\eea
with the notation
\bea
{\tilde a}_{k,l} & = & a_k a_l - q_{lk} a_l a_k \; ,        \label{36}
\\
{\tilde a}_{k,l,m} & = & {\tilde a}_{k,l} a_m -
q_{mk}q_{ml} a_m {\tilde a}_{k,l} \; .                      \label{37}
\eea
Generalizing eqs.(\ref{34}) and (\ref{35}) for arbitrary n,
one arrives at the following relation:
\bea
\left.\sum_{\pi (k,{\bf i}_{n-1})}\right.
\left.\sum_{\sigma (k,{\bf i}_{n-1})}\right.
c_{\pi (k,{\bf i}_{n-1}) , \sigma (k,{\bf i}_{n-1})}
\; a_{\pi (k)}^{\dagger}a_{\pi ( i_{1})}^{\dagger} \cdots
a_{\pi ( i_{n-1})}^{\dagger}
a_{\sigma (k)} a_{\sigma ( i_{1})} \cdots
a_{\sigma ( i_{n-1})} & &
\nn \\
= \; \left.\sum_{\pi ({\bf i}_{n-1})}\right.
\left.\sum_{\sigma ({\bf i}_{n-1})}\right.
( M_n^{-1} )_{k,\sigma ({\bf i}_{n-1}) ;
 k,\pi ({\bf i}_{n-1})}
{\tilde a}_{k,\pi ({\bf i}_{n-1}})^{\dagger}
{\tilde a}_{k, \sigma ({\bf i}_{n-1})} \; , & &             \label{38}
\eea
where
\be
{\tilde a}_{k, j_{1}, \dots , j_{n-1}}=
{\tilde a}_{k, j_{1}, \dots , j_{n-2}} a_{j_{n-1}} -
q_{j_{n-1}k}q_{j_{n-1} j_1} \cdots q_{j_{n-1} j_{n-2}}
a_{j_{n-1}} {\tilde a}_{k, j_{1}, \dots , j_{n-2}} \; .     \label{39}
\ee
Apart from the elegance of eqs. (\ref{34}),(\ref{35}) and
(\ref{38}), the expression of the number operator
makes it easier to relate our results to the results of
Greenberg \cite{gr2}, corresponding to the $q_{ij}=0$
$(\forall i,j)$ statistics.

All the results obtained above correspond to the case when the
deformation parameters satisfy the condition
\be
\mid q_{ij} \mid < 1 \; \; ( \forall i,j ).                 \label{41}
\ee

Before concluding, we briefly discuss the cases where we have
the following conditions, instead of (\ref{41}):
\be
\mid q_{ij} \mid > 1 \; \; ( \forall i,j )                  \label{42}
\ee
or
\be
\mid q_{ij} \mid = 1 \; , \rm {i.e.} \; \;
q_{ij}=e^{i{\Phi}_{ij}}\;\;(\forall i,j ).                  \label{43}
\ee

If the case (\ref{42}) is realized, the number operator exists.
This is clearly seen from eqs. (\ref{14}) and (\ref{16}).
However, as it is evident
from eq.(\ref{15}), the norm of the state vector $ a_{i_{1}}^{\dagger}
a_{i_{2}}^{\dagger} \cdots a_{i_{n}}^{\dagger} \mid 0 >$ is not
positive,
making this case physically unacceptable.

However, if $q_{ij}$'s are such that the condition (\ref{43})
is satisfied, even
though the matrix $M_n(q)$ is singular, the existence of the number
opearator is not excluded. In this case, the existence
of the operator $N_k$
depends on whether or not it is possible to impose such
a $q-$mutator on the
operator pairs $a_i ,a_j~( \forall i,j )$
that all terms higher than $a_k^{\dagger}a_k$ cancel. This is
precisely what happens for $q_{ij}=\pm 1$, corresponding to Bose and
Fermi statistics, respectively. Further investigation regarding
this point is unquestionably of interest and is the subject of
another study. For the statistics characterized
by eqs.(\ref{43}), it
should be pointed out that anyons (particles existing in 2+1
dimensions)
can alternatively be represented as a $q-$deformation of an underlying
bosonic algebra. This can be viewed as an extension of Greenberg's
approach with $q$ being a complex number $\mid q \mid$=1 \cite {dor}.

To conclude, in this paper we have studied the generalized
$q-$deformation of the Heisenberg algebra
defined by eqs. (\ref{3}), (\ref{4}) and (\ref{5}).
For the case when deformation parameters are such that
$\mid q_{ij} \mid <1 (\forall i,j) $
we have proved the existence and presented a method to explicitly
construct the number operator for particles
obeying the corresponding statistics. We have also proved
the positivity of the
norm of linearly independent state vectors.

\newpage
{\large \bf Acknowledgment} \\

I express my gratitude to S.Pallua and S. Meljanac for introducing
me to the subject of q$-$deformed algebras. I wish to thank
V. Bardek for stimulating discussions and comments,
and B. Ni\v zi\' c for careful reading of the manuscript.
I would also like to thank the other members of the Theoretical
Physics Department of the Rudjer Bo\v skovi\' c Institute
for their encouragement and support during the course of
this work.

This work was supported by the Croatian Ministry of Science and
Technology under contract No. 1-03-199 .

\end{document}